# A DATA-DRIVEN APPROACH FOR TURBULENCE MODELING

**Yangmo Zhu, Nam Dinh**
Department of Nuclear Engineering
North Carolina State University, Raleigh, NC 27695
yangmo_zhu@ncsu.edu ; ntdinh@ncsu.edu

**ABSTRACT**

Data-driven turbulence modeling is a newly emerged research area in thermal hydraulics simulation of nuclear power plant (NPP). The most common CFD method used in NPP thermal hydraulics simulation is Reynolds-averaged Navier–Stokes (RANS) method, which still has acknowledged deficiencies not only in the calculation speed but also in the complexity of choosing turbulence model and parameters for different flow patterns. Data-driven turbulence modeling aims to develop a RANS-based method which not only computationally efficient but also applicable to different flow patterns. To achieve this goal, the first step is to develop an approach to properly perform RANS for selected flow patterns. In this work, a machine learning approach is selected to achieve this goal.

The main purpose of this study is to perform a data-driven approach to model turbulence Reynolds stress leveraging the potential of massive direct numerical simulation (DNS) data. The approach is validated by a turbulence flow validation case: a parallel plane quasi-steady state turbulence flow case.

The work contains three parts. The first part is database preparation. In this step, turbulence properties (Reynolds stress) are extracted from DNS results, which are considered as "physically correct data". Meanwhile, flow features are extracted from RANS results, which are considered as "data to be corrected". The second part is surrogate model establishment. In this step, a data-driven regression function is trained between flow features and turbulence properties obtained from the previous step. The last part is model validation, which is applying trained data-driven regression function to a test case to validate this approach.

## 1. INTRODUCTION

Thermal hydraulics simulation, especially turbulence flow simulation of NPP usually cost a lot of computational resources. In research area to evaluate NPP safety, such as probabilistic risk assessment (PRA), hundreds or thousands of simulations are required to perform a safety analysis. In such situation, RANS method is still computationally expensive. Coarse mesh CFD is a newly emerged research area that aims to develop much faster simulation than RANS. But the foundation of it is a universally applicable turbulence model for RANS which enable RANS to properly perform for every flow pattern in NPP. The aim of this work is to discuss a data-driven approach to achieve that goal.

Reynolds-averaged Navier–Stokes (RANS) equations are commonly used in simulating turbulence flow in NPP. The RANS equations are time-averaged equations for fluid flow. The idea behind RANS is Reynolds decomposition, which is a mathematical technique to separate the average and fluctuating parts of a quantity. For a stationary, incompressible Newtonian fluid, these equations can be written in Einstein notation as:

$$\rho \bar{u}_j \frac{\partial \bar{u}_i}{\partial x_j} = \rho \bar{f}_i + \frac{\partial}{\partial x_j}\left[-\bar{p}\delta_{ij} + \mu\left(\frac{\partial \bar{u}_i}{\partial x_j} + \frac{\partial \bar{u}_j}{\partial x_i}\right) - \rho \overline{u'_i u'_j}\right] \qquad (1)$$

The nonlinear Reynolds stress term on the right-hand side of this equation requires additional models to close the RANS equations and hence led to the creation of many different turbulence models. Actually, it is the appearance of Reynolds stress that makes the prediction of RANS usually inaccurate compare to the reality. On the one hand, the simple idea to provide extra equations (turbulence model) usually do not work. And even a turbulence model could work for a particular class of flow, it will most likely not be able to work in even a slightly different environment. On the other hand, compiling engineering tables for design handbooks could also bring substantial risk. Even when based on a wealth of experience, a sufficient number of validation tests are required to see if the tables can be extrapolated to a particular situation, which is often too expensive to perform [1].

In view of the development of data science these years, people starts to pursue solving this problem through machine learning methods [2-5]. By definition, machine learning refers to a process of using data to build regression functions of responses with respect to input variables. The trained functions can be evaluated by predicting test cases where data are not included in the training dataset. Despite the diversity in achieving final goal, the aim of turbulence modeling for different approaches are the same: to improve the predictive capability of turbulence models. However, there is still no consensus on the choice of input and response for the machine learning algorithm. Eric and Duraisamy [2] introduce a full-field discrepancy factor $\beta$ as the learning response, while Xiao et al. [4] choose the discrepancy between Reynolds stress in RANS model and DNS as learning response for the reason that they're trying to model the model-form uncertainties generated from RANS-model. Although both the $\beta$ and Reynolds stress discrepancy are demonstrated to be able to recover true result, $\beta$ is modeled quantities and have a relatively less physical interpretation, Reynolds stress discrepancy is highly related to the Reynolds stress of RANS model, which could be an inaccurate model by itself. As a result, the trained model could only be applied to the same RANS model with fixed model form and parameters. For the input features, Duraisamy [2] used full-field non-dimensional flow and model variables to construct the input features. However, the full-field input feature is restricted by the field geometry and boundary conditions, thus trained functions for a certain case is hard to be applied to different geometry or boundary conditions. Ling and Templeton [5] point out the importance of embedding invariance properties into machine learning process. In this study, we choose high-fidelity Reynolds stress from DNS as model response and low-fidelity flow feature from RANS as model input to train the model. The reason for such treatment is based on the practice that Reynolds stress shows better performance in propagating mean velocities. Such treatment is theoretically equivalent to model the discrepancy between Reynolds stress in RANS model and DNS, as the Reynolds stress of RANS model could be derived from input flow feature.

In addition, there is also a significant barrier between predicting Reynolds stress and propagating velocities. Here the term "propagation" is used to refer the procedure of obtaining flow solution from RANS equations using predicted Reynolds stresses[4]. A number of challenges exist when propagating the corrected Reynolds stresses through RANS equations to obtain the mean velocity and pressure fields. On the one hand, the error introduced by the difference of mesh size between high-fidelity simulation and low-fidelity simulation could propagate during this process. On the other hand, the nonlinearity of the equation also increases along with Reynolds number, which further increases the difficulty in converging. The objective of this study is to introduce the approach to predict Reynolds stress based on high fidelity (HF) DNS database and low fidelity RANS result as a baseline. Moreover, the study also demonstrates its capability of improving both Reynolds stress and propagated averaged velocities in a relative simple benchmark with DNS data.

## 2. METHODOLOGY

This section summarizes the framework of data-driven turbulence modeling. Its key procedures and components are discussed, which includes preparation of the input flow feature and output responses of machine learning model, and building of regression functions.

This work aims to demonstrate the framework of data-driven turbulence modeling. Specifically, given Reynolds stresses from high-fidelity DNS data and flow features from low-fidelity RANS data of training flows, a data-driven model would be trained to predict the Reynolds stress for different flows without DNS data. Here, training flow refers to the flow used to train the machine learning model, such training data are DNS or experimental data that could be considered as "physically correct" data. Accordingly, test flow is the flow to be used to validate the model.

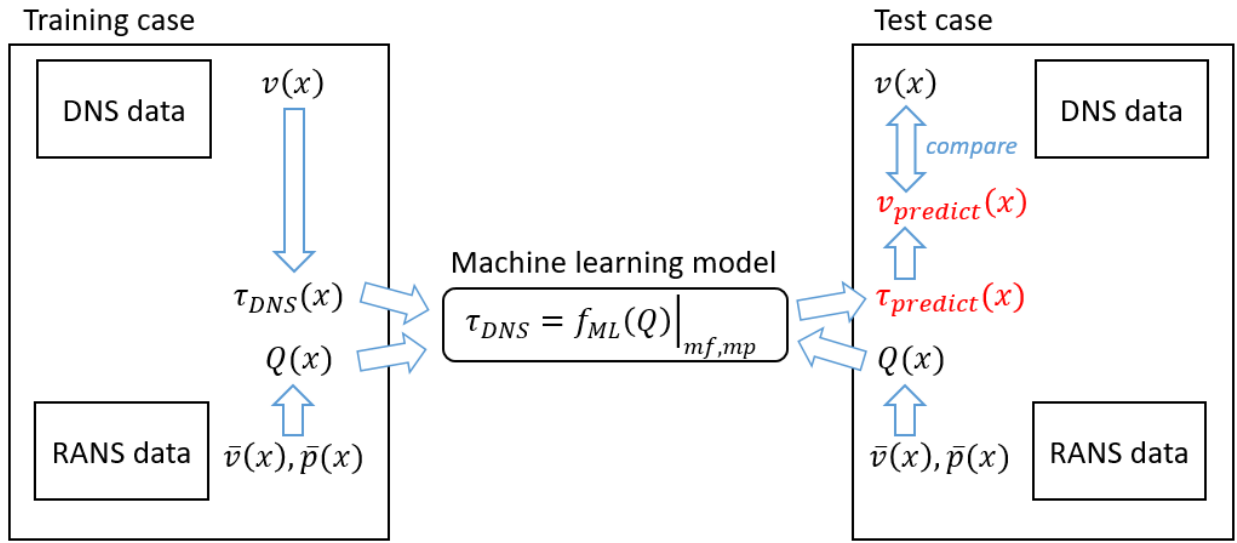

**Figure 1. The framework flowchart**

The overall procedure can be summarized as follow:
1. Database preparation. In this step, turbulence properties (Reynolds stresses) $\tau$ are extracted from DNS results, which are considered as output response of machine learning model. Flow features $Q$ are extracted from RANS results, which are considered as input for machine learning model.
2. Surrogate model establishment. In this step, a data-driven regression function $f_{ML}$ is trained between flow feature $Q$ and turbulence properties $\tau$ with machine learning algorithm.
3. Model validation. In this step, the trained data-driven regression function is taken to predict Reynolds stresses $\tau$ of a test case. Then, the predicted Reynolds stresses are taken to the RANS equations to propagated velocity field. After that, the DNS velocity fields are used to validate the model.

Theoretically, the machine learning regression function is a function of many variables:

$$\tau = f_{ML}(Q, mf, mp, \Delta_{\text{DNS}}, \Delta_{\text{RANS}}) \qquad (2)$$

Here we make the first assumption that there is a 1 vs 1 mapping relationship between the high-fidelity DNS Reynolds stresses and a set of model inputs, which include low-fidelity flow feature $Q$, model form

$mf$ and model parameter $mp$ of RANS turbulence model, mesh sizes $\Delta$ of DNS and RANS. According to this equation, the uncertainty of model response $d\tau$ could be expressed as:

$$d\tau = \left(\frac{\partial \tau}{\partial Q}\right)_{mf,mp,\Delta_{DNS},\Delta_{RANS}} dQ, \left(\frac{\partial \tau}{\partial mf}\right)_{Q,mp,\Delta_{DNS},\Delta_{RANS}} dmf, \left(\frac{\partial \tau}{\partial mp}\right)_{Q,mf,\Delta_{DNS},\Delta_{RANS}} dmp,$$
$$\left(\frac{\partial \tau}{\partial \Delta_{DNS}}\right)_{Q,mf,mp,\Delta_{RANS}} d\Delta_{DNS}, \left(\frac{\partial \tau}{\partial \Delta_{RANS}}\right)_{Q,mf,mp,\Delta_{DNS}} d\Delta_{RANS}, \quad (3)$$
$$df_{ML}(Q, mf, mp, \Delta_{DNS}, \Delta_{RANS})$$

As can be seen, the uncertainty of model prediction result comes from 6 sources: input flow features uncertainty, RANS model uncertainties (include model form uncertainty and model parameter uncertainty), high-fidelity data uncertainty, discretization uncertainty and machine learning algorithm uncertainty. The relationship between these 6 terms is nonlinear and difficult to derive, hence we only separate them by commas.

In order to simplify the study, fixed model form and parameters are applied to both training flow and test flow so that the second term and third term on the right-hand side of equation (3) could be neglected. We further assume that high-fidelity data uncertainty, discretization uncertainty, and machine learning algorithm uncertainty is small enough to be negligible. Hence the uncertainty of model prediction result is only dominated by the input flow feature $Q$. Finally, equation (2) could be rewritten as:

$$\tau = f_{ML}(Q)|_{mf,mp} \quad (4)$$

In equation (4), Reynolds stresses are only related to input flow feature $Q$. This is the final equation applied in Fig. 1.

## 2.1. Input and response of machine learning model

One of the most critical parts of machine learning algorithm is the selection of input and response. First, model inputs should generally reveal the main feature of local flow; then, inputs should be normalized quantities so that the trained model could be applied to more general cases. Moreover, the input feature should be nondirectional in order to filter superfluous data. According to these requirements, a systematic methodology of constructing a complete invariant input set from a group of given tensorial variables as suggested by Ling et al. [7] is employed in the current work.

$$Q^+ = \{S^+, R^+, \nabla p^+\} \quad (5)$$

Here we assume that the local flow feature $Q$ could be described by strain rate $S$, rotation rate $R$, and pressure gradient $\nabla p$, which are also widely used in traditional turbulence modeling. Xiao et al. [6] also take the gradient of turbulence kinetic energy $\nabla k$ as one of the inputs. The reason it is not taken in this work is that we consider that flow with same strain rate $S$ and rotation rate $R$ should have the same gradient of turbulence kinetic energy $\nabla k$ given the same RANS model. So that $\nabla k$ is eliminated in order to improve the machine learning process. All input parameters in this study are normalized in wall units by friction velocity $u_\tau$ and viscosity $\nu$ with a superscript "+". For example, dimensionless wall distance $y^+$ is equal to $u_\tau y/\nu$.

Mean Reynolds stress $\tau^+$ obtained from high-fidelity (DNS) data is selected as model response. Since the major source of model-form errors in RANS simulation comes from modeled Reynolds stress, it is a natural choice to directly learn the Reynolds stresses from DNS data. Although some may doubt that such treatment would abandon RANS model and solely rely on data instead, it could be derived that the choice of Reynolds stress is equivalent to the discrepancy of Reynolds stress between DNS and RANS, cause Reynolds stress of RANS could be calculated from RANS mode and $S, R$ and $\nabla p$. Here we select DNS Reynolds stress as response because in practice DNS Reynolds stress shows better performance in propagating mean velocities.

## 2.2. Construction of Machine Learning Model

After data of the model input and response being prepared, a machine learning algorithm is constructed using Gaussian process. There are various choices of machine learning algorithm, for example, neural network [3], random forest [6]. Gaussian has 2 advantages that lead it to be used in this work. First, the Gaussian process provides confidence range and prediction distribution of the prediction result based on data sufficiency, which could largely help decision making; Second, hyperparameters in the Gaussian process could be self-optimized, which eliminate human error in selecting model parameters.

It should be noticed that data refinement is needed in the Gaussian process. As the Gaussian process is highly depended on the density of data, the predicting result would obviously bias to the area where data are more sufficient, even those data are mostly overlapped or too close to provide meaningful information. Hence data refinement is required to rearrange data into an evenly distributed structure.

## 3. NUMERICAL RESULTS

### 3.1. Case Setup

Cases of fully developed incompressible pressure-driven turbulent flow between two parallel planes are performed in this work. Periodic boundary conditions are applied in the streamwise (x) and spanwise (z) directions, and no-slip boundary conditions are applied to the wall. The computational domain sizes are 8pi meters in streamwise domain and 3pi in the spanwise domain.

**Table I. Parameters setup for cases (SI Units)**

|  | Case 1 | Case 2 | Case 3 | Case 4 | Case 5 |
|---|---|---|---|---|---|
| **Density ($\rho$)** | 1 | 1 | 1 | 1 | 1 |
| **Kinematic Viscosity ($\nu$)** | 0.00035 | 0.0001 | 0.00005 | 0.000023 | 0.000008 |
| **Channel half width ($\delta$)** | 1 | 1 | 1 | 1 | 1 |
| **Avg streamwise velelocity ($\bar{u}$)** | 1 | 1 | 1 | 1 | 1 |
| **Friction velocity ($u_\tau$)** | 6.373e-02 | 5.434e-02 | 5.002e-02 | 0.04587 | 0.04148 |
| **Shear Reynold's number ($Re_\tau$)** | 182.088 | 543.496 | 1000.512 | 1994.756 | 5185.897 |
| **Bulk Reynold's number ($Re_b$)** | 2857 | 10000 | 20000 | 43650 | 125000 |
| **Wall shear stress ($\tau_w$)** | 4.06e-03 | 2.95e-03 | 2.50e-03 | 2.10e-03 | 1.72e-03 |
| **RANS Mesh number in wall normal direction** | 6 | 18 | 33 | 66 | 172 |

As can be seen in Table I, totally 5 cases are performed in 5 different Reynolds numbers. All flows have the same geometry. Data of these flows are obtained from DNS simulation [8]. Case 2,3,5 are selected as training data and case 4 is selected as test data. Training cases and test case are differentiated by Re number. Data of case 1 is abandoned due to the potential numerical error introduced by the limit number of mesh size.

Baseline RANS simulations are performed for each flow to obtain input flow feature for machine learning model. Standard k-epsilon model is selected with following parameters: $C_\mu = 0.09, C_1 = 1.44, C_2 =$

$1.92, \sigma_\epsilon = 1.3$. An open-source CFD software, OpenFoam, is used to perform this study with a built-in incompressible flow solver pimpleFoam. Spatially even distributed mesh structure is applied and mesh sizes for these cases are presented in Table I. In order to save calculation expense, high-Reynolds number k-epsilon model with wall function treatment is used instead of low-Reynolds number k-epsilon model. In order to apply wall function to the calculation domain, the node that is nearest to the wall should be in log-law region, which means $y^+$ of it should be larger than 30. The RANS mesh number in wall normal direction is calculated based on this criterion. In this work, it is assumed that the error introduced by the difference of Von Karman constants in each case could be ignored, hence one wall function is applied to all the cases with standard Von Karman constants to be $\kappa = 0.41, E = 9.8$. Future work may change from high-Reynolds number k-epsilon model to low-Reynolds number k-epsilon model so that the limitation of mesh size could be overcome and numerical discretization error could be evaluated. Correlation-based machine-learning model for wall function constants will also be studied.

The Gaussian process is performed using GPML code written by Carl Edward Rasmussen and Hannes Nickisch [9]. The isotropic squared exponential covariance function "covSEiso" and Gaussian likelihood function "likGauss" are selected to be likelihood function. Hyperparameters are self-optimized subsequently. Based on our testing, a maximum number of 93 function evaluations is large enough to have a robust prediction.

### 3.2. Results

### 3.2.1. Model Validation

Since the aim of this framework is to improve modeling of turbulence Reynolds stress. It is a natural choice to validate the model by comparing velocity field propagated from machine learning model and that of DNS data. Theoretically, one should be able to obtain an accurate prediction of the velocity field with the predicted Reynolds stresses. However, the outcome depends on a lot of factors. Although Reynolds stresses from DNS simulations are supposed to be closer to the DNS Reynolds stresses than those from RANS predictions, there is no guarantee that a better velocity field could be propagated due to the potential convergence risk. According to Thompson et al. [10], even for channel flows, the Reynolds stresses of different DNS databases in literature lead to significant discrepancies in the propagated velocity fields. Hence, in order to evaluate the performance of machine learning model, it is better to start with a comparison of velocity field generated by machine learning model and DNS data. Here case 4 is selected as a test case, and case 2, 3 and 5 are selected as training cases to implement machine learning model to predict Reynold stresses in case 4.

As can be seen in Fig.2, the DNS velocity profile is plotted along with the propagated mean velocity profile for comparison. It can be seen that the predicted results agree well with the DNS profiles in the log-law region, the error between prediction result and DNS result is very small. Even though, an error analysis is always recommended to be performed. In this study, possible sources of error come from 4 aspects: the first one is HF data error, it should be validated that the Reynolds stress calculated by DNS could be taken back to RANS equations and make the residuals equal to zero. This type of error should be taken special care of when process DNS data; the second one is wall function error, only 1 wall function is applied in this study. The difference between this wall function and DNS data could introduce error to the final result; the third one is machine learning model prediction error. The differences of predicted Reynolds stress and DNS Reynolds stress are shown in the following sections. The last source of error is discretization error, which exists in almost every CFD experiment. Nevertheless, the overall data quality of Reynolds stresses is considered satisfactory to obtain an improved velocity field.

To analysis the HF data error. Reynolds stress should be taken back to RANS equations to calculate residuals of each equation. As this is a quasi-steady state problem. The RANS equation could be simplified as equations (6-7):

$$\rho \frac{\partial \overline{u'v'}}{\partial y} - \frac{\partial p}{\partial x} = \mu \frac{\partial u^2}{\partial y^2} \tag{6}$$

$$\frac{\partial p}{\partial y} = \rho \frac{\partial \overline{v'v'}}{\partial y} \tag{7}$$

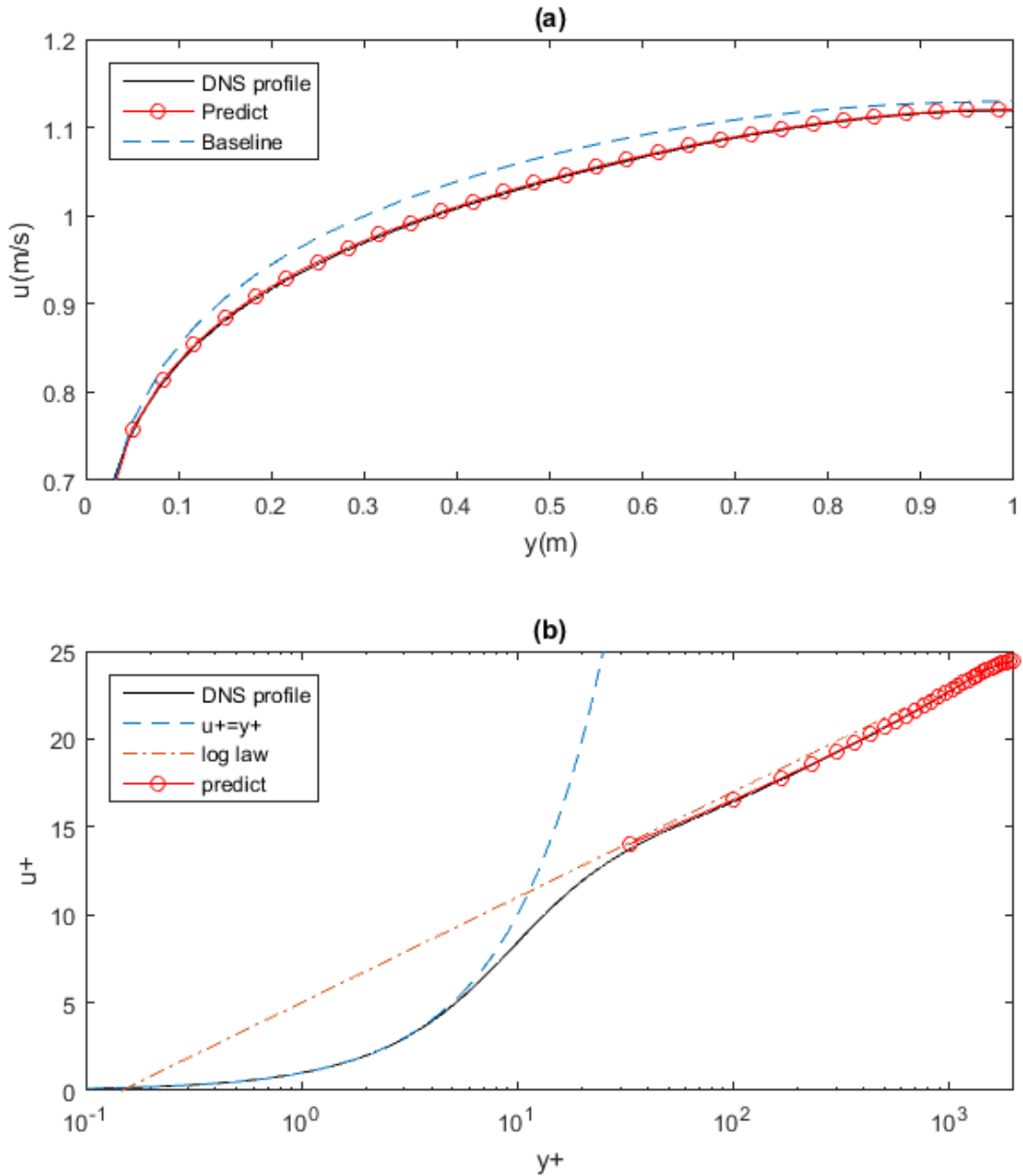

Figure 2. Comparison of predicted and DNS velocity profile via normal direction of the plane. Normal profile (a), normalized log axis profile (b)

According to the simplified RANS equations. Residuals could be calculated as fig. 3.

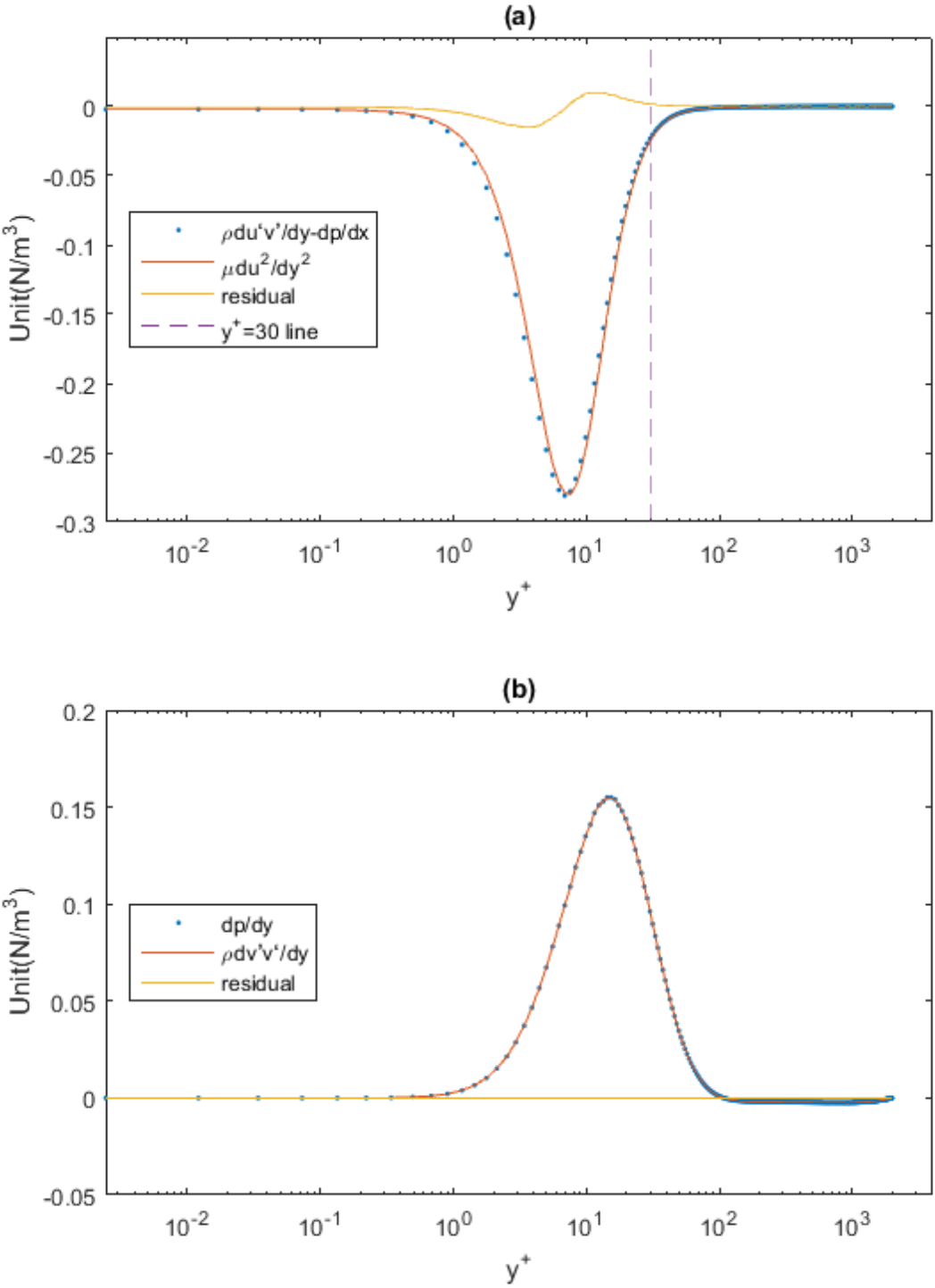

**Figure 3. Residuals of equation 6 (a) and 7 (b) in y-direction.**

As can be seen in fig. 3, the residuals of equation 7 are equal to 0. But the residuals of equation 6 aren't. Such non-zero residuals could possibly be introduced by the different discretization schemes applied in DNS and RANS, or just due to discretization error. Nevertheless, these non-zero residuals only show in the region where $y^+ < 30$. As wall function treatment is applied in this work, only residuals in the region where $y^+ > 30$ could effect simulation result. Hence here we assume HF data error is very small and could be ignored.

### 3.2.2. Learning and Prediction of Reynolds stress

The normalized Reynolds stresses $<uu>^+, <vv>^+, <ww>^+, <uv>^+$ are learned from training flow, and predictions are made for test flow. The prediction results are plotted in fig. 4.

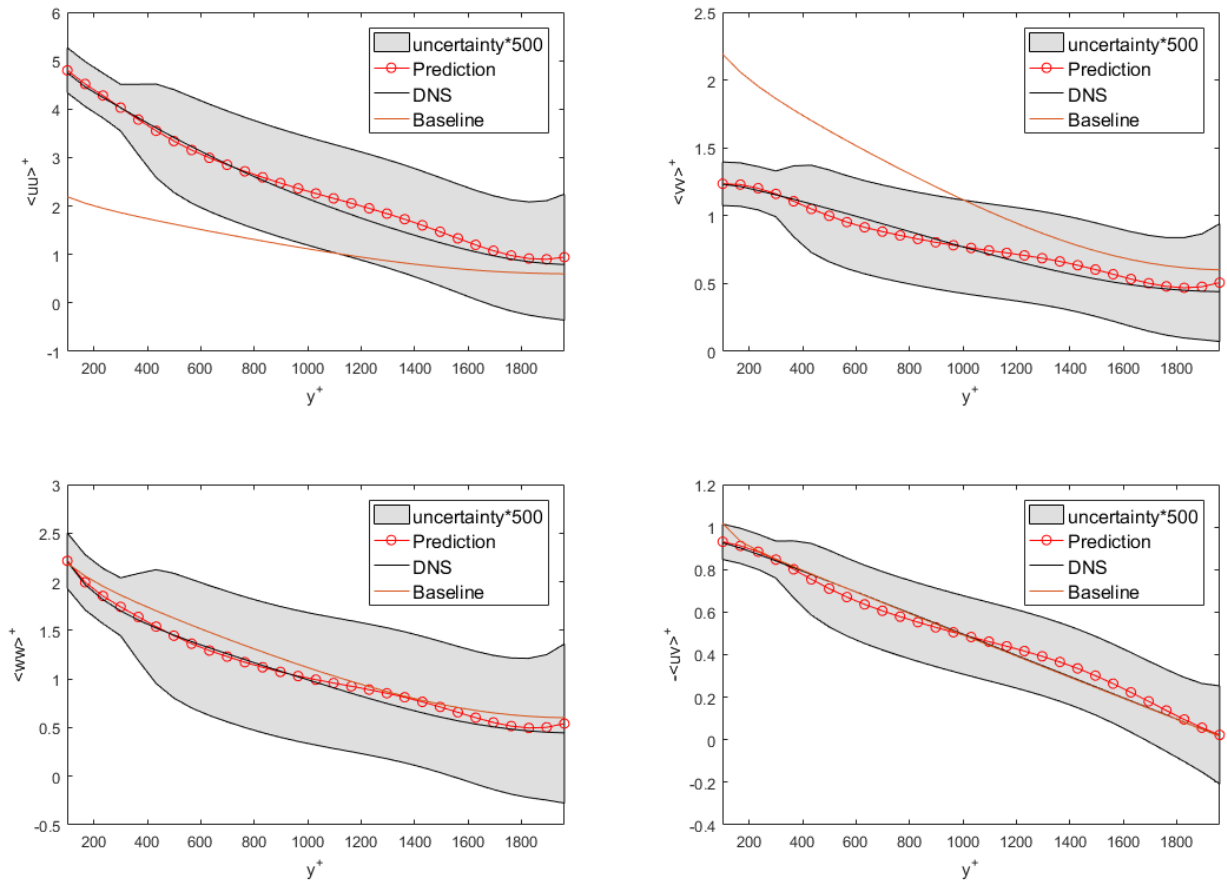

**Figure 4. A comparison of normalized Reynolds stress via normal direction of the plane in DNS, RANS baseline and predicted result (along with uncertainty range amplified by 500 for the convenience of viewing).**

In this figure, we plot a comparison of normalized Reynolds stress for the predicted result, DNS result, and RANS baseline result. In order to show the capability of the data-driven model to provide a degree of confidence in its predictions, we also plot uncertainty ranges in the figure. The uncertainty ranges are multiplied by 500 for the convenience of reading.

As can be seen in fig. 4, the predicted result of the 4 stresses are all close to the DNS result but with little difference. While the RANS baseline result of $<uv>^+$ is already very close to the DNS result. The uncertainty ranges are small in the near wall region and grow larger as $y^+$ increases. For example, in the prediction of $<uv>^+$, the RANS baseline result is different from DNS in the near-wall region, but machine learning prediction compensates this error. The reason uncertainty grows along with $y^+$ is because there are more data in low $y^+$ region. Training data include case 2,3 and 5. As the Reynolds number increase with case number, the data distribution extends to much higher $y^+$ value.

It could be seen from fig. 4 that RANS predicted $<uu>^+, <vv>^+$ have large discrepancy from DNS data. But such difference does not cause too much difference between the result of RANS and DNS. The reason is because it is the combination of $<uu>^+, <vv>^+$ and $<ww>^+$, the turbulence kinetic energy $k^+$, that exist in the RANS k-epsilon model. As can be seen in fig. 4, the turbulence kinetic energy $k^+$ of RANS doesn't vary too much from DNS result.

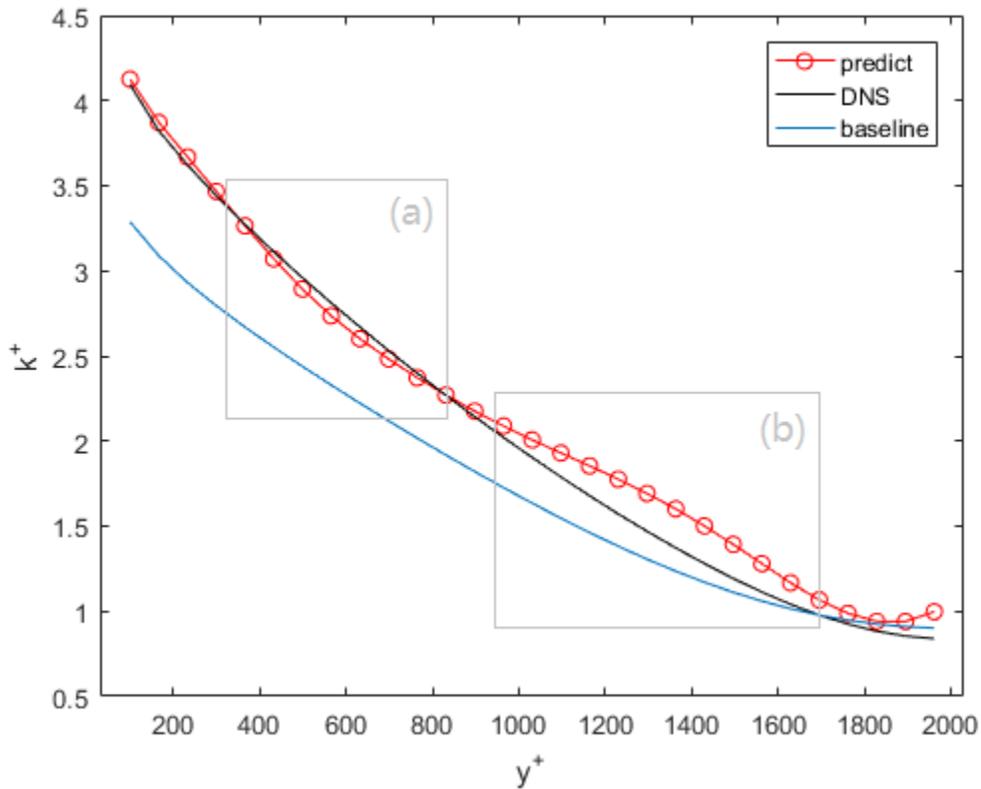

**Figure 5. A comparison of normalized turbulence kinetic energy via normal direction of the plane in predicted result, DNS and RANS baseline.**

As can be seen from fig. 4 & 5, the machine learning model tends to underestimate Reynolds stress in the relative near wall region (a) and overestimate Reynolds stress in remote wall region (b). Such error is because of the characteristic of the Gaussian process. The Gaussian process bias to the area where data are more sufficient. As shown in fig. 6, case 2,3,4,5 cover $y^+$ range of [0,151], [0,513], [0,967], [0,1961] and [0,5155], separately. Hence region (b) are only covered by data from case 4 and 5, in which case 4 is test data. So, the dominant training data in region (b) is from case 5. In this situation, Gaussian process prediction will bias to the case 5 result, which is higher than case 4, as can be seen in fig. 6 (a). A similar phenomenon could be seen in region (a), where dominant data are not case 5 but case 2,3, which are all relatively low Reynolds stress cases compared with case 4. Hence Gaussian process underestimates Reynolds stress in this region.

The gradient of streamwise velocity in wall normal direction $\frac{du^+}{dy^+}$ is an important component of input flow feature $Q$. Hence the relationship between it, wall normalized distance $y^+$ and turbulence kinetic energy $k^+$ for the 5 cases are plotted in fig. 6. Here we first obtain Reynolds stresses $\tau$ from DNS data as a function of position, as fig. 6 (a); Then, we obtain input flow feature $Q$ from RANS data as a function of position, as fig. 6 (c). After that, mapping functions between Reynolds stress and local flow feature are established using machine learning algorithm, as fig. 6 (b).

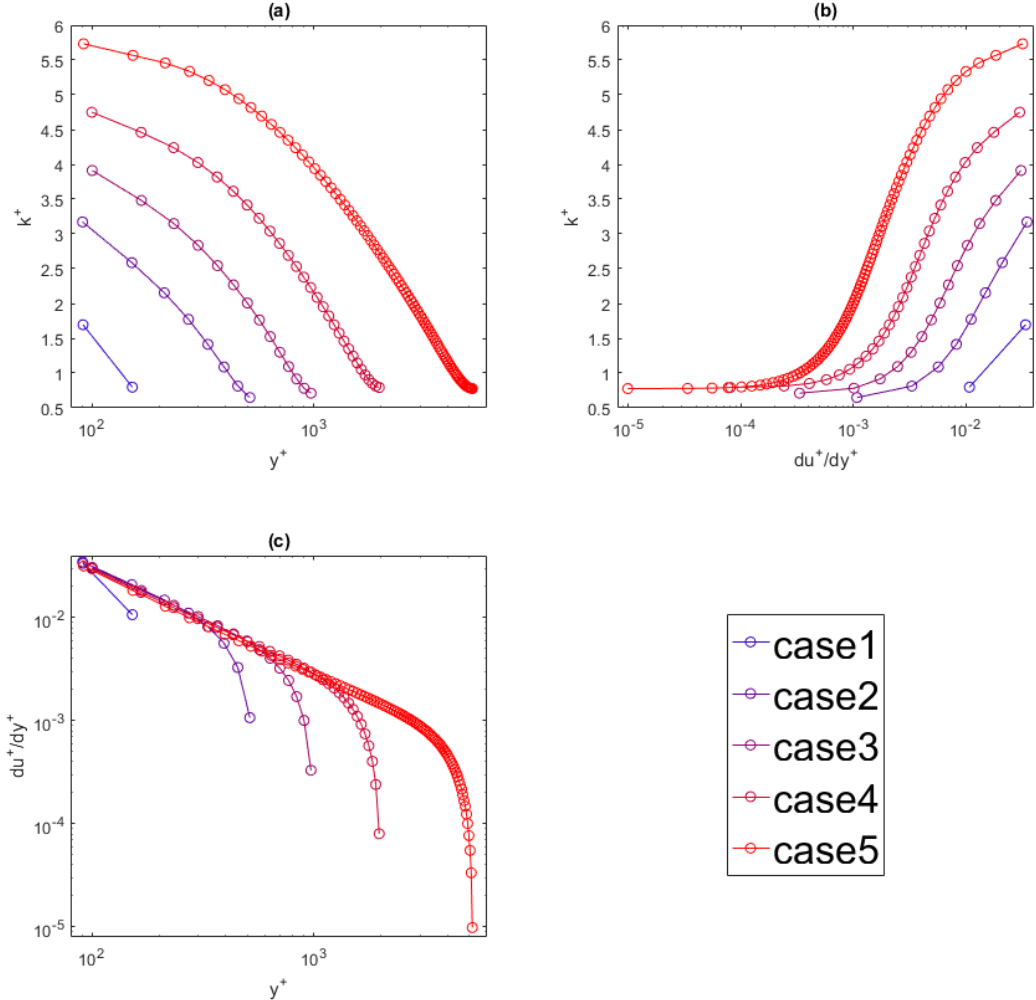

**Figure 6.** The relationship between $k^+$, $\frac{du^+}{dy^+}$ and $y^+$ in 5 cases

## 4. DISCUSSION AND FUTURE WORK

In data-driven turbulence modeling, uncertainties come from 4 parts: model uncertainty (include model form uncertainty and model parameter uncertainty), high-fidelity data uncertainty, discretization uncertainty and machine learning algorithm uncertainty, as listed in Table II. This work focuses on modeling the model form uncertainty so that the modeling of model parameter uncertainty, high-fidelity data uncertainty, and discretization uncertainty are not performed. In future research, model parameters would also be studied as it is also an important part of the model uncertainty. The influence model parameters could do to Reynolds stress are much more complicated than that of Reynolds number, which brings us a bigger challenge.

Although the process works well for this parallel plate benchmark, there are still challenges prevent this research from going future. First of them is the propagation problem. We have also attempted to establish mapping functions between DNS Reynolds stress and DNS flow feature, then take the functions into RANS to propagate velocity field, which is a physically more reasonable process. The biggest problem that stops us is the difficulty of propagation. Due to the non-linearity of RANS equation in high Re number case, even little prediction error could be amplified and lead to divergence of the propagation. How to properly propagate predicted Reynolds stress into RANS equation is an inevitable topic in future research. Another challenge is the selection of high-fidelity data. Currently, we use DNS data as HF data because only DNS result provide unmodeled Reynolds stress. But for many engineering problems, HF data may be RANS, LES or even experiment data. How to apply these HF data into the data-driven model is another challenge in future research.

**Table II. Sources of error in data-driven turbulence modeling**

| Source of error | Comments |
|---|---|
| **Model form uncertainty** | Data-driven turbulence modeling requires a RANS baseline to provide input flow feature, hence the selection of turbulence model (k-epsilon, k-omega, etc.) could introduce uncertainty to the final result. In this work, the standard k-epsilon is selected to perform all cases so that the model form uncertainty between case 1~5 should be equal. |
| **Model parameter uncertainty** | Similar to model form uncertainty, different turbulence model has different model parameters, and usually, such parameters are remained to be calibrated for different scenarios. Hence the selection of model parameter could also introduce uncertainty. |
| **High-fidelity data uncertainty** | The data-driven method requires high-fidelity data to drive the whole model, hence it is critical to validate that the high-fidelity data is relevant and sufficiently accurate. Inaccurate data or their extrapolation would result in inaccurate prediction. |
| **Discretization uncertainty** | Discretization uncertainty exists in the process of performing RANS baseline, machine learning modeling, and velocity propagation. It is difficult and usually unnecessary to model discretization uncertainty. Control this uncertainty in an acceptable range is the main research direction. |
| **Machine learning algorithm uncertainty** | Different machine learning algorithm has difference feature; hence the choice of machine learning algorithm could also introduce uncertainty. Moreover, the machine learning process could even contain human error (choose network layers, model parameters, etc.) or random error (in some machine learning algorithm like Random Forest method). |

After a reliable data-driven model been developed, our next goal is to model the discretization uncertainty in it. The modeling discretization uncertainty is critical for coarse mesh CFD in NPP. Once the discretization uncertainty becomes predictable, much faster modeling could be expected for industrial applications.

## 5. CONCLUSIONS

In this study, we introduce a data-driven approach to predict Reynolds stresses of flows in a new application using pre-calculated HF data. According to the case study result, the data-driven model largely improved RANS model. The improved RANS model allows more accurate and flexible prediction for turbulence flow, which is not only the aim of turbulence modeling but also fundamental for coarse mesh CFD in NPP. As nuclear safety problem becomes increasingly stringent and requires detailed analysis of thermal-hydraulics, computationally efficient CFD methods are needed to support the study of a broad range of accident scenarios. Current turbulence modeling requires different RANS models for different flow patterns, which largely hinder the use of CFD for complex scenarios with transient flow patterns. The aim of solving the problem using data-driven method is the starting point of this study. Basically, there are still several challenges: propagation problem, machine learning model bias, and HF data error. Future work will focus on more complex cases (complex geometry and multi-phase problem) to evaluate the challenges and explore different propositions to address the challenges.

## ACKNOWLEDGMENTS

The support for this work by the U.S. Department of Energy through the Nuclear Energy University Program (Integrated Research Project) and the Idaho National Laboratory's National University Consortium program is gratefully acknowledged.